\documentclass[aps,prb,twocolumn]{revtex4-1}
\usepackage{amsfonts}
\usepackage{amsmath}
\usepackage{amssymb}
\usepackage[final,dvips]{graphicx}
\begin{document}
\title{Statistics of subgap states in $s_\pm$ superconductors}
\author{A. Glatz and A. E. Koshelev}
\affiliation{Materials Science Division, Argonne National Laboratory, Argonne,
Illinois 60439}

\pacs{74.20.-z,74.62.En,74.70.Xa}
\date{\today}

\begin{abstract}
There is strong support in favor of an unusual $s_{\pm}$ superconducting state
in the new iron-based superconductors, in which the gap parameter has opposite
signs in different bands. In this case scattering between different bands by
impurities has a pair-breaking effect and introduces states inside the gap. We
studied the statistics of disorder-induced subgap states in $s_{\pm}$
superconductors due to collective effects of impurities. Numerically solving
the two-band Bogolyubov equations, we explored the behavior of the density of
states and localization length. We located the mobility edge separating the
localized and delocalized states for the 3D case and the crossover between the
weak and strong localization regimes for the 2D case. We found that the widely
used self-consistent T-matrix approximation is not very accurate in describing
subgap states.
\end{abstract}
\maketitle

\begin{figure}[ptb]
\begin{center}
\includegraphics[width=2.8in]{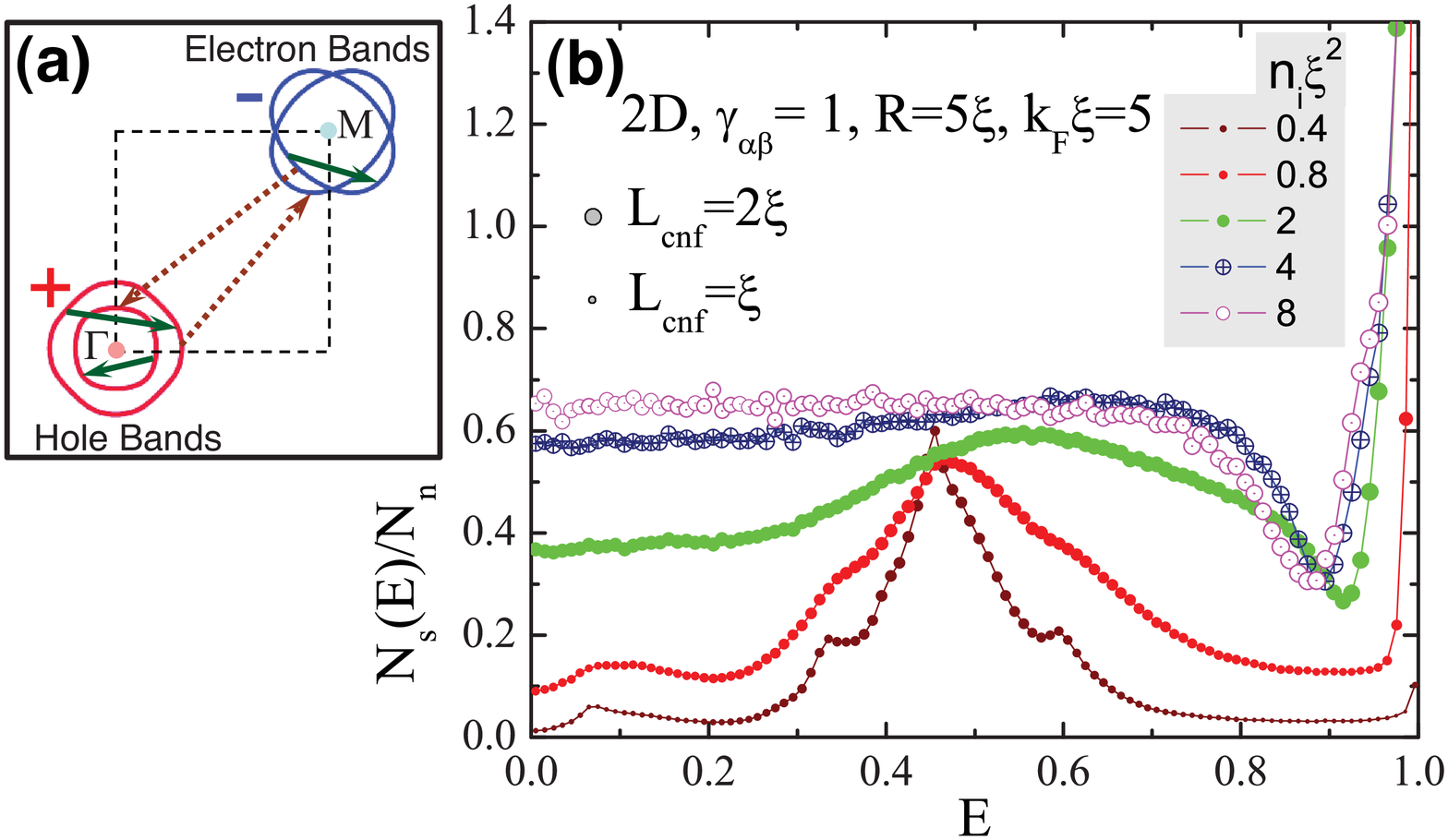}
\end{center}
\caption{(color online)(a)The schematic band structure of iron pnictides. Solid
and dotted arrows illustrate intraband and interband scattering events.
(b)Evolution of the 2D subgap DoS with increasing concentration of impurities
$n_{i}$ for isotropic scattering. Symbol sizes are proportional to
$L_{\mathrm{cnf}}$ (\ref{LocLengthNum}). } \label{Fig-DoSConc_Imp}
\end{figure}

The recent discovery of high-temperature superconductivity in the iron arsenide
LaFeAsO$_{1-x}$F$_{x}$ \cite{Kamihara} followed by the discovery of several new
classes of superconducting materials \cite{IronPnictides} led to a major
breakthrough in the field of superconductivity. The Fermi surface of these
materials is composed of several electron and hole sheets located near
different points of the Brillouin zone \cite{BandStruct}, see Fig.\
\ref{Fig-DoSConc_Imp}a. There is theoretical reasoning in favor of an
electronic origin of superconductivity and a unusual superconducting state in
which the order parameter has opposite signs in different bands (s$_{\pm}$
state).\cite{spm-state} This scenario is supported by the observation of a
resonant magnetic mode in the superconducting state by inelastic neutron
scattering.\cite{ChristiansonNat08}

In the s$_{\pm}$-state interband scattering due to potential impurities (see
Fig.\ \ref{Fig-DoSConc_Imp}a) has a pair-breaking effect similar to magnetic
impurities in conventional superconductors.\cite{AGZhETF60,ShibaPTP68,RusinovPZheTF68} Such
impurities introduce states inside the gap known as Shiba-Rusinov states for
the magnetic-impurities problem.\cite{ShibaPTP68,RusinovPZheTF68,BalatskyRMP06} This leads to
a finite density of states (DoS) at zero energy and strongly influences the
superconducting properties at low temperatures. Most iron pnictides are doped
superconductors and disorder due to dopant atoms is unavoidable. Experimental
properties of iron pnictides which may be explained by a finite DoS at the
Fermi level due to pair-breaking disorder include the ``residual'' linear
temperature dependence of the specific heat in the superconducting state
\cite{SpecHeat} and a quadratic temperature dependence of the London
penetration depth at low temperatures. \cite{LonPenT2}

Transport properties at low temperatures, such as thermal conductivity
\cite{ThermCond} and microwave surface resistance, are sensitive to
localization state of the low-energy quasiparticles. A random impurity
potential may localize quasiparticles within some energy range. For 3D
superconductors we typically expect two regimes: (i) At small concentrations of
impurities states at the Fermi level are localized and a mobility edge  at a
finite energy exists, similar to impurity bands in semiconductors; (ii) At
sufficiently high impurity concentrations all states are delocalized. The
critical impurity concentration depends on the scattering properties of the
impurities. In the first regime, the localized states near zero energy
contribute to the low-temperature behavior of the thermodynamic but not
transport properties.

A standard analytical approach to describe collective effects of impurities in
superconductors is the self-consistent T-matrix approximation (SCTM). For the
Born limit, it was elaborated in the famous paper by Abrikosov and
Gor'kov\cite{AGZhETF60} and was later generalized for strong scattering
\cite{ShibaPTP68}.  Most theoretical works on collective effects of
impurities in superconductors  are based on this approximation, see, e.g., the
review [\onlinecite{BalatskyRMP06}]. While giving a qualitative description of the
impurity band, the SCTM approximation has serious deficiencies. For small
impurity concentrations it predicts a hard gap in the spectrum. In reality, the
DoS is finite at all energies, it has an exponential tail\cite{LifshitzTail}
due to rare fluctuation configurations of impurities, similar to the Lifshitz
tail in the impurity band of semiconductors. Another deficiency of the SCTM
description is that it ignores localization properties of the states.
Localization of quasiparticles in superconductors was studied for dirty d-wave
superconductors \cite{LocQPdwave} and for the mixed state of disordered s-wave
superconductors \cite{LocQPvortex}. To our knowledge, localization in the
impurity band of superconductors with pair-breaking impurities was never
studied.

Several recent papers address different aspects of the impurity-induced subgap
states \cite{PreostiMuzPRB96,SengaJPSJ08,Matsumoto09} and their possible
influence on properties of iron pnictides \cite{VorontsovPRB09}. Studies of
collective impurity effects, however, do not go beyond the SCTM approach.
Motivated by the importance of disorder-induced subgap states for the
properties of iron pnictides and the absence of an accurate theoretical
description of these states, we performed a detailed study of their statistical
properties based on numerical calculations. We explore the behavior of the
density of states for s$_{\pm}$ superconductors as function of scattering
parameters and impurity concentration and compare the results with the SCTM
approach. We also explore localization properties of states in order to locate
the mobility edge in the parameter space.

Our study is based on the two-band Bogolyubov equations for the two-component
wave functions, $\hat{\Psi}_{\alpha}=\binom{u_{\alpha} }{v_{\alpha}}$,
\[
\left(  E\!-\!\hat{\varepsilon}_{\alpha}\hat{\tau}_{z}\!+\!\Delta_{\alpha}
\hat{\tau}_{x}\right)  \hat{\Psi}_{\alpha}(\mathbf{r})-\hat{\tau}_{z}
\sum_{l,\beta}\delta(\mathbf{r}-\mathbf{R}_{l})U_{\alpha\beta}^{l}\hat{\Psi
}_{\beta}(\mathbf{r})\!=\!0.
\]
Here $\alpha$ is the band index, $\hat{\tau}_{i}$ are Pauli matrices in Nambu
space, $\hat{\varepsilon}_{\alpha}=\xi_{\alpha
}(\hat{\mathbf{k}})-\varepsilon_{F}\approx\mathbf{v}_{F,\alpha}(\hat
{\mathbf{k}}-\mathbf{k}_{F,\alpha})$,  $\Delta_{\alpha}$ are the gap
parameters. We assume that $\Delta_2=-\Delta_1$. The last term in the equation
describes the interaction with impurities. The interband scattering is
described by the off-diagonal terms in the matrix $U_{\alpha\beta}^{l}$. For
scattering between bands $1$ and $2$, separated by wave vector $\mathbf{Q}$
equal to half of the reciprocal-lattice vector, $U_{12}^{l}$ contains the
factor $\exp(i\mathbf{QR}_{l})$ which only takes values $\pm 1$ depending on
$\mathbf{R}_{l}$. This means that even for identical impurities $U_{12}^{l}$
has random signs and its average is zero. We neglect inhomogeneities of the gap
parameters due to impurities. It is known that these inhomogeneities are small
and do not influence the quasiparticle states much. The key parameter of an
isolated pair-breaking impurity is the energy of a localized
state\cite{Matsumoto09}, $E_{0}/\Delta\equiv
\varepsilon_{0}=\sqrt{1-4\Gamma_{\mathrm{eff}}}$,
 where we introduced the effective interband scattering parameter,
\begin{equation}
\Gamma_{\mathrm{eff}}\!=\!\frac{\gamma_{12}\gamma_{21}}{1\!+\!\gamma_{22}
^{2}\!+\!\gamma_{11}^{2}\!+\!2\gamma_{12}\gamma_{21}\!+\!\left(  \gamma
_{22}\gamma_{11}\!-\!\gamma_{12}\gamma_{21}\right)  ^{2}},\label{EffInterband}
\end{equation}
with $\gamma_{\alpha\beta}=\pi\nu_{\alpha}U_{\alpha\beta}$ being the reduced
scattering amplitudes and $\nu_{\alpha}\!=\!\int\!\frac
{d^{d}\mathbf{k}}{(2\pi)^{d}}\delta(\xi_{\alpha}(\mathbf{k})\!-\!\varepsilon_{F})
$ being the normal DoS (per spin) for band $\alpha$.

We explore the properties of the subgap states for the two- and
three-dimensional cases. For the numerical analysis, we rewrite the equations
in a form containing only wave functions at the impurity sites,
\begin{equation}
\hat{\Psi}_{\alpha}(\mathbf{R}_{l})=\sum_{\beta,l^{\prime}}\hat{g}_{\alpha
}(\mathbf{R}_{l}-\mathbf{R}_{l^{\prime}})\hat{\tau}_{z}\gamma_{\alpha\beta
}^{l^{\prime}}\hat{\Psi}_{\beta}(\mathbf{R}_{l^{\prime}}),\label{EqImpSites1}
\end{equation}
where the reduced  Green's function is defined as
\[
\hat{g}_{\alpha}(\mathbf{R})=\frac{1}{\pi\nu_{\alpha}}\int\frac
{d^{d}\mathbf{k}}{(2\pi)^{d}}\exp(i\mathbf{kR})\frac{E+\varepsilon_{\alpha
}(\mathbf{k})\hat{\tau}_{z}-\Delta_{\alpha}\hat{\tau}_{x}}{E^{2}
-\varepsilon_{\alpha}^{2}(\mathbf{k})-\Delta_{\alpha}^{2}}
\]
and $d$ is the spacial dimensionality.
For each impurity realization inside a box of size $R^d$ we find the set of
eigenenergies $E_\lambda$ and the corresponding four-component wave functions
$\hat{\Psi}_l^{\lambda}\!=\!
[\hat{\Psi}_{1}^{\lambda}(\mathbf{R}_{l}),\hat{\Psi}_{2}^{\lambda}(\mathbf{R}_{l})]$,
see appendix~\ref{app.num}. From the set of eigenenergies,
we compute the average DoS, $N_{s}(E)=\left\langle
\sum_{\lambda}\delta(E-E_{\lambda})\right\rangle$, where the average is taken
over many impurity realizations.  We normalize $N_{s}(E)$ to the total normal
DoS for excitations, $N_{n}=4\nu$. To characterize localization properties, we
compute the average confinement length for given energy,
\begin{equation}
L_{\mathrm{cnf}}(E,R)=\left\langle \sqrt{\sum_{a=x,y[,z]} \left(\left\langle
r_{a}^{2}\right\rangle _{\lambda} -\left\langle r_{a}\right\rangle
_{\lambda}^{2}\right) } \right\rangle _{E_{\lambda}=E} ,\label{LocLengthNum}
\end{equation}
where $\left\langle r_{a}^m\right\rangle _{\lambda} \!
=\!\sum_{l}R_{l,a}^m|\Psi _{l}^{\lambda}|^{2}$ ($m\!=\!1,2$), and
$|\Psi_{l}^{\lambda}|^{2}\!=\!\sum_{\alpha}\left[  |u_{\alpha}^{\lambda
}(\mathbf{R}_{l})|^{2}\!+\!|v_{\alpha}^{\lambda}(\mathbf{R}_{l})|^{2}\right]$.
The behavior of $L_{\mathrm{cnf}}(E,R)$ with increasing system size, $R$,
determines the nature of the states. For delocalized states
$L_{\mathrm{cnf}}(E,R)$ is limited by the system size and linearly grows with
$R$. For localized states $L_{\mathrm{cnf}}(E,R)$ saturates at a finite value,
which gives the average localization length of states with energy $E$, $
\lim_{R\rightarrow\infty}L_{\mathrm{cnf}}(E,R)=L_{\mathrm{loc}}(E) $.

We study the subgap densities of states at different concentrations of
impurities and scattering parameters.  We consider first the case of isotropic
scattering, when all matrix elements $\gamma_{\alpha\beta}$ are equal. Figure
\ref{Fig-DoSConc_Imp}b shows the evolution of the subgap DoS with increasing
concentration of impurities $n_{i}$ for a moderate scattering rate,
$\gamma_{\alpha\beta}=1$ for all $\alpha$ and $\beta$, and the system size
$R=5\xi$. The energy is normalized to the gap value $\Delta$ and the coherence
length is defined as $\xi= v_{F}/\Delta$. For small concentrations of
impurities the DoS has a peak near the energy of the localized state. With
increasing impurity concentrations, the peak broadens and becomes completely
smeared already at relatively small impurity concentration $n_{i}\xi^{2}=2$. At
higher concentrations the DoS becomes almost constant and comparable with the
normal DoS.
\begin{figure}[ptb]
\begin{center}
\includegraphics[width=3.0in]{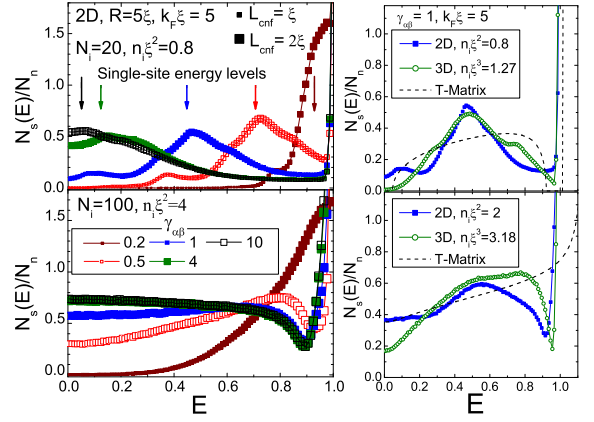}
\end{center}
\caption{(color online) \emph{Left: }Evolution of the subgap DoS with changing
scattering rate for isotropic scattering for two impurity concentrations.
Symbol sizes are proportional to the confinement length.
 \emph{Right: }Comparison of the DoS shapes for the 2D and 3D models with
identical pair-breaking parameters and SCTM results for two impurity concentrations.}
\label{Fig-DoS_2D3D}
\end{figure}
The left part of Fig. \ref{Fig-DoS_2D3D} shows the subgap DoS for two impurity
concentrations for a wide range of isotropic scattering rates. The energy
location of the peak for small concentrations goes down with increasing
scattering strength, while the maximum DoS is almost independent on
$\gamma_{\alpha \beta}$. The peak smears with increasing impurity concentration
and the magnitude of the density of states at large concentrations depends only
weakly on the scattering strength for $\gamma_{\alpha\beta}>0.5$. Also, for
large scattering rates the DoS approaches a limiting shape corresponding to the
unitary limit. For small scattering strength, $\gamma_{\alpha\beta}=0.2$,  we
found the typical ``Lifshitz tail'' behavior.

Within the SCTM approximation  (see appendix~\ref{app.SCTM}) the DoS does not depend on
dimensionality of superconductor, it is determined by the location of the
single-site energy level and the pair-breaking parameter proportional to the
impurity concentration $\alpha= 2n_{i}\Gamma_{\mathrm{eff}}/(\pi\nu\Delta)$.
Even though the overall evolution of the DoS shape with variations of the
scattering strength and impurity concentration is similar in the 2D and 3D
cases, the universality suggested by the SCTM approach does not exist.  In the
right part of Fig.\ \ref{Fig-DoS_2D3D} we compare the computed DoS for two
impurity concentrations corresponding to the  same pair-breaking strength in
the 2D and 3D cases with the SCTM results. One can see that the DoS shapes for
the two dimensionalities are similar but not identical. One noticeable
difference is that the 3D DoS is typically smaller at low energies. The SCTM
approximation does not reproduce the DoS shape at low impurity concentrations,
the sharp peak in the center and the small features at the sides are not
reproduced. These features appear due to the oscillating dependence of the
impurity pair energy on their separation and correspond to the pairs separated
by the distance at which the energies have extrema (see appendix~\ref{app.pairs}). For
large impurity concentration we found a pronounced dip in the DoS for energies
slightly smaller than $\Delta$, also not reproduced by the SCTM approximation.

\begin{figure}[ptb]
\begin{center}
\includegraphics[width=3.4in]{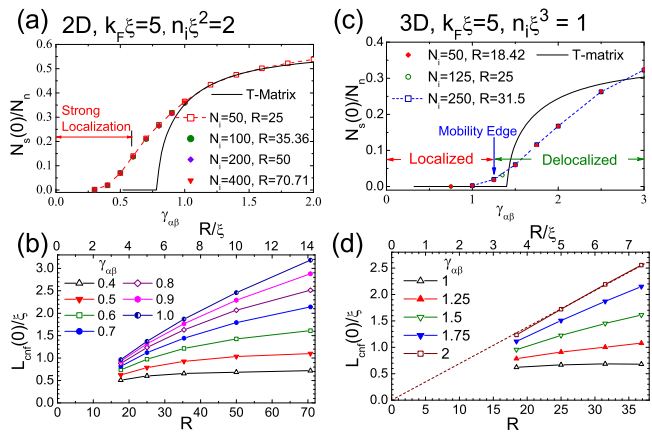}
\end{center}
\caption{(color online) Plots (a) and (c) show the dependences of the
zero-energy DoS on the scattering strength for isotropic scattering  for 2D and
3D cases. Solid lines show the SCTM results.  Plots (b) and (d) show the size
dependences of the confinement length at different scattering strength.}
\label{Fig-DoS_Lcnf_Ni_2D3D}
\end{figure}
We now focus on the region near zero energy, which determines the
low-temperature behavior of the superconducting parameters. Figures
\ref{Fig-DoS_Lcnf_Ni_2D3D}(a,c) show the dependences of the zero-energy DoS on
the isotropic scattering strength for fixed impurity concentration for the 2D
and 3D models. We see that the DoS has a negligible size effect even at the
smallest studied sizes. The SCTM approximation only roughly reproduces the the
shape of the numerical curves and does not describe the tail region. The size
dependences of the confinement length are shown in Fig.\
\ref{Fig-DoS_Lcnf_Ni_2D3D}(b,d). We can see that for the 2D case at
$\gamma_{\alpha\beta}< 0.6$ the confinement length saturates at large $R$
approaching a finite localization length. No clear saturation of
$L_{\mathrm{cnf} }$ is observed for $\gamma_{\alpha\beta}> 0.6$. This may imply
that states are delocalized or the localization length may be much larger than
the studied system sizes. The second possibility looks more plausible, because
for 2D disordered systems all electronic states are expected to be localized.
The region $\gamma_{\alpha\beta}\approx0.6$ probably marks a sharp crossover
between the weak and strong localization regimes. We actually observe a
noticeable downward curvature in the dependences of $L_{\mathrm{cnf}}$ vs. $R$
for all $\gamma_{\alpha\beta}$ which may be interpreted as a tendency towards
localization at larger length scales. For the 3D case we expect a true mobility
edge which we can estimate as the value of $\gamma_{\alpha\beta}$ at which
$L_{\mathrm{cnf}}$ has a clear saturation tendency at large $R$. The estimated
location of the mobility edge, $\gamma_{\alpha\beta}\approx 1.25$, is marked in
Fig.\ \ref{Fig-DoS_Lcnf_Ni_2D3D}c. It is close to the critical value bounding
the SCTM gapped regions. We found that the  DoS values at the mobility edge are
quite small, ($\sim 0.02$ in our example). These values are significantly
smaller than the DoS at the localization crossover for the 2D case. This
observation implies that for identical bands in the 3D case there is only a
very narrow parameter window within which the localized states at zero energy
provide a noticeable DoS.

\begin{figure}[ptb]
\begin{center}
\includegraphics[width=3.2in]{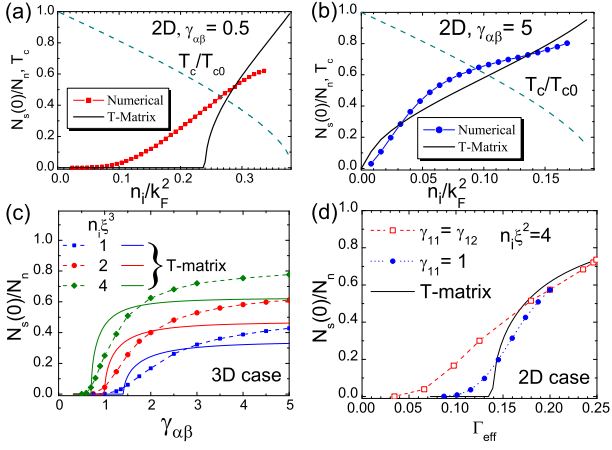}
\end{center}
\caption{(color online) Plots (a) and (b) show the dependences of the
zero-energy 2D DoS and transition temperature on concentration of impurities
for two isotropic scattering rates. The clean-limit coherence length is taken
as $k_{F}\xi\!=\!2.63$. (c)The dependences of the zero-energy 3D DoS on the
scattering strength for three concentrations of impurities. (d)Densities of
states as function of the interband scattering parameter
$\Gamma_{\mathrm{eff}}$, Eq.\ (\ref{EffInterband}), for two simulations series,
isotropic scattering, $\gamma_{11}=\gamma_{12}$, and fixed intraband scattering
amplitude, $\gamma_{11}=1$. } \label{Fig-N0Tc-ni}
\end{figure}
In figure \ref{Fig-N0Tc-ni}(a,b) we compare the numerical and SCTM dependences
of the zero-energy 2D DoS on the impurities concentration for weak and strong
isotropic scattering. We also show the concentration dependences of the
transition temperature evaluated using the Abrikosov-Gor'kov
formula (see Ref.~[\onlinecite{AGZhETF60}] and appendix~\ref{app.SCTM}). For small scattering strength a
noticeable DoS appears at the Fermi level only when $T_{c}$ is strongly
suppressed. In contrast, for large scattering strength the DoS reaches values
comparable with the normal DoS already at very minor suppression of $T_{c}$.

In Fig.\ \ref{Fig-N0Tc-ni}c we present the dependences of the zero-energy DoS
on the scattering strength for three concentrations of impurities,
$n_{i}\xi^{3}=1$, $2$, and $4$ and compare them with the predictions of the
SCTM approximation, which only coarsely reproduces the evolution of the DoS
with increasing concentration. It does not describe the tail regions and
systematically underestimates the value of the zero-energy DoS in the unitary
limit, corresponding to large $\gamma_{\alpha\beta}$.

Up to this point we considered the case of isotropic scattering when all
scattering amplitudes are equal. In general, one can expect that the interband
scattering is always weaker than the intraband one. Within the SCTM approach
the DoS is not sensitive to the individual scattering amplitudes and is
completely determined by the parameter $\Gamma_{\mathrm{eff}}$, Eq.\
(\ref{EffInterband}). To clarify the role of the relative strength of the
scattering amplitudes, we study the subgap states for different interband
amplitudes $\gamma_{12}=\gamma_{21}$ at fixed intraband amplitudes
$\gamma_{11}=\gamma_{22}$. The DoS dependences on $\Gamma_{\mathrm{eff}}$ for
the two simulation series are compared in Fig.\ \ref{Fig-N0Tc-ni}d. We can see
that the DoS is not simply determined by $\Gamma_{\mathrm{eff}}$, as suggested
by the SCTM approach, but also sensitive to the relative strength of the
intraband and interband scattering. For the same $\Gamma_{\mathrm{eff}}$ the
DoS decreases with decreasing ratio $\gamma_{12}/\gamma_{11}$.

In conclusion, we explored the subgap DoS and localization properties for
disordered s$_{\pm}$ superconductors. We found that the widely-used analytical
description (SCTM) is incomplete and not very accurate. Disorder makes
superconductivity ``gapless'', the DoS at $E=0$ is always finite. In the 3D
case there is a mobility edge separating localized and delocalized states. It
reaches zero energy at a critical impurity concentration above which all states
become delocalized. In the 2D case the mobility edge is replaced by a crossover
separating strongly and weakly localized states. The development of
quantitative theory of the subgap states is crucial for the understanding
properties of the iron pnictides and other superconductors with pair-breaking
impurities.

We would like to thank I. Vekhter, T. Proslier, K. Matveev, and U. Welp for
useful discussions. This work is supported by UChicago Argonne, LLC, under
contract No. DE-AC02-06CH11357 and by the Center for Emergent
Superconductivity, an Energy Frontier Research Center funded by the U.S. DOE,
Office of Science, Office of BES under Award No. DE-AC0298CH1088

\appendix

\section{Self-Consistent T-matrix approximation}\label{app.SCTM}

The self-consistent T-matrix approximation\cite{ShibaPTP68} is defined by the
coupled equations for the average Green's function $\hat{G}_{\alpha\beta
}(\mathbf{k})\equiv\delta_{\alpha\beta}\hat{G}_{\alpha}(\mathbf{k})$ and the
self-energy part $\hat{\Sigma}_{\alpha}\equiv\delta_{\alpha\beta}\hat{\Sigma
}_{\alpha\alpha}$,
\begin{subequations}
\label{SCTM}
\begin{align}
&  \hat{G}_{\alpha}(\mathbf{k})=\left[  E-\hat{\varepsilon}_{\alpha}\hat{\tau
}_{z}+\Delta_{\alpha}\hat{\tau}_{x}-\hat{\Sigma}_{\alpha}\right]  ^{-1},\\
&  \hat{\Sigma}_{\alpha}=n_{i}\hat{T}_{\alpha}=n_{i}\left\langle \left(
1-\hat{\tau}_{z}U_{\alpha\beta}\hat{G}_{0,\beta}\right)  ^{-1}\hat{\tau}
_{z}U_{\beta\alpha}\right\rangle
\end{align}
with $\hat{G}_{0,\alpha}=\int\frac{d^{d}\mathbf{k}}{(2\pi)^{d}}\hat{G}
_{\alpha}(\mathbf{k})$. Note that the off-diagonal components of $\hat{\Sigma
}_{\alpha\beta}$ average to zero due to the random signs of the off-diagonal
components of $U_{\alpha\beta}$. The partial density of states is related to
the Green's function as $N_{\alpha}(E)=\operatorname{Im}\left\{
\mathrm{Tr}\left[  \hat{G}_{0,\alpha}(E-i\delta)\right]  \right\}  $. Using the
expansion $\hat{\Sigma}_{\alpha}=\sum_{i}\Sigma_{\alpha,i}\hat{\tau}_{i}$ with
$i=0,x,y,z$, we can express $N_{\alpha}(E)$ via the ratio $u_{\alpha}=\left(
E-\Sigma_{\alpha,0}\right)  /\left( \Delta-\Sigma_{\alpha,x}\right)  $ as
\end{subequations}
\begin{equation}
N_{\alpha}(E)=2\nu_{\alpha}\operatorname{Im}\frac{u_{\alpha}}{\sqrt
{1-u_{\alpha}^{2}}}.\label{DoS}
\end{equation}
Here the factor $2$ accounts for the electron and hole excitations.

In the case of identical bands, this system reduces to an equation for only one
complex parameter $u=\left(  E-\Sigma_{0}\right)  /\left(  \Delta-\Sigma
_{x}\right)  $, similar to the Shiba equation for magnetic
impurities\cite{ShibaPTP68},
\begin{align}
&  u\left(  1-\alpha\frac{\sqrt{1-u^{2}}}{\varepsilon_{0}^{2}-u^{2}}\right)
=\frac{E}{\Delta},\label{ShibaEq}\\
&  \alpha=\frac{2n_{i}\Gamma_{\mathrm{eff}}}{\pi\nu\Delta}=\frac{1}{\tau
_{12}\Delta}\\
&  =\frac{2n_{i}}{\pi\nu\Delta}\frac{\gamma_{12}\gamma_{21}}{1+\gamma_{22}
^{2}+\gamma_{11}^{2}+2\gamma_{12}\gamma_{21}+\left(  \gamma_{22}\gamma
_{11}-\gamma_{12}\gamma_{21}\right)  ^{2}}.\nonumber
\end{align}
We observe that increasing the intraband scattering potential $\gamma
_{\alpha\alpha}$ increases the intraband scattering time $\tau_{12}$ and
diminishes the pair-breaking parameter $\alpha$.\cite{SengaJPSJ08} We also note
that the SCTM results do not depend on dimensionality of the superconductor.

We summarize the most important analytical results obtained within this
approximation. The SCTM approximation gives a gapped state for $\alpha
<\varepsilon_{0}^{2}$ and gapless state for $\alpha>\varepsilon_{0}^{2}$ with
finite total density of state at zero energy
\begin{equation}
N_{s}(0)=4\nu\frac{\sqrt{\alpha^{2}-\varepsilon_{0}^{4}}}{\sqrt{\frac
{\alpha^{2} }{2}+\varepsilon_{0}^{2}\left(  1-\varepsilon_{0}^{2}\right)
+\alpha\sqrt{\frac{\alpha^{2}}{4}+1-\varepsilon_{0}^{2}}}}.\label{N0Tmatrix}
\end{equation}
We stress again that the existence of a \emph{gapped} state is unrealistic
feature and deficiency of this approximation.

The suppression of $T_{c}$ is determined by the Abrikosov-Gor'kov formula
\cite{AGZhETF60},
\begin{equation}
\ln\frac{T_{c0}}{T_{c}} =\psi(1/2+1/2\pi\tau_{12} T_{c})-\psi
(1/2),\label{SupprTc}
\end{equation}
where $\psi(x)$ is the digamma function. This famous result is almost always
understood too literally. In fact, it just gives an approximate typical value
of the critical temperature. In real systems, due to the random arrangement of
impurities, the transition temperature is inhomogeneous and the transition has
percolative nature.

The average zero-temperature gap parameter is determined by the equations
\cite{RusinovPZheTF68}
\begin{subequations}
\begin{equation}
\ln\frac{\Delta_{0}}{\Delta}=\frac{\pi}{2}\frac{\alpha}{1+\varepsilon_{0}
}\text{ for }\alpha<\varepsilon_{0}^{2}\label{GapSCTM1}
\end{equation}
with $\alpha=\alpha_{0}\Delta_{0}/\Delta$ and
\begin{align}
\ln\frac{\Delta_{0}}{\Delta} &  =\frac{\pi}{2}\frac{\alpha}{\varepsilon_{0}
+1}+\ln\left(u_{0}+\sqrt{1+u_{0}^{2}}\right)
-\frac{\alpha u_{0}}{u_{0}^{2}\!+\!\varepsilon_{0}^{2}}
\nonumber\\
&  \!+\!\frac{\alpha }{\varepsilon_{0}^{2}\!-\!1}\left[  \arctan
u_{0}\!-\!\varepsilon_{0}
\arctan\left(  \frac{u_{0}}{\varepsilon_{0}}\right)  \right]  ,\label{GapSCTM2}\\
\text{with }u_{0}^{2}  =&\frac{\alpha^{2}}{2}-\varepsilon_{0}^{2}+\alpha
\sqrt{\frac{\alpha^{2}}{4}+1-\varepsilon_{0}^{2}}, \text{ for }
\alpha>\varepsilon_{0}^{2}\nonumber
\end{align}
\end{subequations}
Here $\Delta_{0}$ is the gap parameter for the clean case.

\section{Numerical simulations}\label{app.num}

To develop a precise theoretical description of the subgap region, we solve
 Eqs.\ (2) of the main paper numerically for the two-dimensional and
three-dimensional cases. The main element of these equations is the Green's
function in real space, $\hat{g}_{\alpha }(r)$. At $r=0$ the Green's function
does not depend on dimensionality,
\[
\hat{g}_{\alpha}(0)=\frac{-E+\Delta_{\alpha}\hat{\tau}_{x}}{\sqrt
{\Delta_{\alpha}^{2}-E^{2}}}.
\]
Large-$r$ asymptotics of $\hat{g}_{\alpha }(r)$ for $k_{F,\alpha}r\gg1$ are
given by
\begin{widetext}
\begin{align*}
\hat{g}_{\alpha }(r)& =\left[ \frac{-E+\Delta _{\alpha }\hat{\tau}_{x}}{
\sqrt{\Delta _{\alpha }^{2}-E^{2}}}\cos \left( k_{F,\alpha }r-\frac{\pi }{4}
\right) -\hat{\tau}_{z}\sin \left( k_{F},_{\alpha }r-\frac{\pi }{4}\right)
\right] \frac{\exp \left( -\sqrt{\Delta _{\alpha }^{2}-E^{2}}r/v_{F,\alpha
}\right) }{\sqrt{\pi k_{F,\alpha }r/2}}\text{, for 2D case} \\
\hat{g}_{\alpha }(r)& =\left[ \frac{-E+\Delta _{\alpha }\hat{\tau}_{x}}{
\sqrt{\Delta _{\alpha }^{2}-E^{2}}}\sin (k_{F,\alpha }r)-\hat{\tau}_{z}\cos
(k_{F,\alpha }r)\right] \frac{\exp (-\sqrt{\Delta _{\alpha }^{2}-E^{2}}
r/v_{F,\alpha})}{k_{F,\alpha }r}\text{,  for 3D case}
\end{align*}
\end{widetext}
As the probability to find two impurities at distance $\sim 1/k_{F}$ is very
small, the structure of states is mostly determined by these asymptotics and
the value at $r=0$. To match the large-$r$ asymptotics with the $r=0$ value, we
use an approximate forms of the Green's functions. For the 2D case we use
\begin{align}
\hat{g}_{\alpha}(r)  &  =\left[  \frac{-E+\Delta_{\alpha}\hat{\tau}_{x}}
{\sqrt{\Delta_{\alpha}^{2}-E^{2}}}J_{0}\left(  k_{F,\alpha}r\right)
+\hat{\tau}_{z}J_{1}\left(  k_{F,\alpha}r\right)  \right] \nonumber \\
&  \times\exp\left(  -\sqrt{\Delta_{\alpha}^{2}-E^{2}}r/v_{F,\alpha}\right)
,\label{ApproxGreens2D}
\end{align}
where $J_{0}\left(  x\right)  $ and $J_{1}\left(  x\right)  $ are Bessel
functions and for the 3D case we use
\begin{align}
\hat{g}_{\alpha}(r)  & \! =\!\left[  \frac{-E\!+\!\Delta_{\alpha}\hat{\tau
}_{x}} {\sqrt{\Delta_{\alpha}^{2}\!-\!E^{2}}}\frac{\sin(k_{F,\alpha}
r)}{k_{F,\alpha} r}-\hat{\tau}_{z} \mathcal{U}(k_{F,\alpha}r)\frac
{\cos(k_{F,\alpha}r)}{k_{F,\alpha}r} \right] \nonumber\\
&  \times\exp(-\sqrt{\Delta_{\alpha}^{2}-E^{2}}r/v_{F,\alpha})
\label{ApproxGreens3D}
\end{align}
with the interpolation function $\mathcal{U}(z)=z^{2}/(z^{2}+1)$. We expect
that the exact behavior of the Green's functions at $k_{F,\alpha}r\!\sim\! 1$
not reproduced by these interpolations have little influence on the properties
of the subgap states. In this paper we limit ourselves with the simplest case
of two equivalent bands, meaning that $k_{F,1}=k_{F,2}\equiv k_{F}$,
$v_{F,1}=v_{F,2}\equiv v_{F}$, and $\Delta_1=-\Delta_2\equiv \Delta$. The gap
$\Delta$ is used as a unit of energy and $k_F^{-1}$ is used as a unit of
length.

We employ the following numerical procedure. First, we define an impurity
realization by $N_{i}$ random coordinates [$\mathbf{R}_{l}$] in a box,
$0<R_{l,a}<R$, and random impurity signs $\delta_{l}=\pm1$ for the off-diagonal
scattering amplitudes, $U_{12}^l=\delta_{l}U_{12}$. From the linear
$4N_{i}\times4N_{i}$ system defined by Eq.\ (2) of the main paper, we find the
eigenenergies, $E_{\lambda}$, and corresponding eigenstates
$\Psi^{\lambda}(\mathbf{R}_{l})$. From the set of eigenenergies, $E_{\lambda}$,
we compute the average density of state
\begin{equation}
N_{s}(E)=\left\langle \sum_{\lambda}\delta(E-E_{\lambda})\right\rangle ,
\end{equation}
where average is taken over many impurity realizations. Practically, this
implies that for every realization we find the number of states $\Delta N(E)$
falling within the energy interval $[E-\Delta E/2,E+\Delta E/2]$ and then
compute the average $N_{s}(E)=\langle\Delta N(E)/(\Delta ER^{2})\rangle$. As an
isolated impurity generates one localized state, for small concentration of
impurities the normalization condition $\int_{0}^{\Delta}N_{s}(E)dE=N_{i}
/R^{2}$ is satisfied. We normalize $N_{s}(E)$ to the total normal density of
states for excitations, $N_{n}=4\nu$, where for the 2D case the single-band DoS
per electron is given by $\nu=k_{F}/(2\pi v_{F})$ and for the 3D case,
$\nu=k_{F}^{2}/(2\pi^{2}v_{F})$.

To characterize localization properties, we also compute the average
confinement length for states at given energy,
\begin{equation}
L_{\mathrm{cnf}}(E,R)=\left\langle \sqrt{\sum_{a}\left\langle \delta r_{a}
^{2}\right\rangle _{\lambda}}\right\rangle _{E_{\lambda}=E}
,
\end{equation}
where $a=x,y[,z]$ and the confinement length of state $\lambda$, $\left\langle
\delta r_{a}^{2}\right\rangle _{\lambda}$, is determined by its wave function
$\Psi_{l}^{\lambda}$ as
\begin{align*}
\left\langle \delta r_{a}^{2}\right\rangle _{\lambda}  &  =\left\langle
r_{a}^{2}\right\rangle _{\lambda}-\left\langle r_{a}\right\rangle _{\lambda
}^{2},\\
\left\langle r_{a}\right\rangle _{\lambda}  &  =\sum_{l}R_{l,a}|\Psi
_{l}^{\lambda}|^{2},\ \ \left\langle r_{a}^{2}\right\rangle _{\lambda}
=\sum_{l}R_{l,a}^{2}|\Psi_{l}^{\lambda}|^{2}
\end{align*}
with $|\Psi_{l}^{\lambda}|^{2}\!=\!\sum_{\alpha}\left[  |u_{\alpha}^{\lambda
}(\mathbf{R}_{l})|^{2}\!+\!|v_{\alpha}^{\lambda}(\mathbf{R}_{l})|^{2}\right] $
and $\sum_{l}|\Psi_{l}^{\lambda}|^{2}\!=\!1$. The behavior of the confinement
length with increasing system size, $R$, determines wether states at given
energy are localized or not. For delocalized states, $L_{\mathrm{cnf}}(E,R)$ is
limited by the system size and grows proportionally to $R$. For localized
states, $L_{\mathrm{cnf}}(E,R)$ saturates at a finite value, which gives the
average localization length of states with energy $E$, $
\lim_{R\rightarrow\infty}L_{\mathrm{cnf}}(E,R)=L_{\mathrm{loc}}(E). $ In three
dimensions for relatively small concentration of impurities one can expect the
existence of a mobility edge in the subgap region separating localized and
delocalized states. It is defined as the energy $E_{\mathrm{ME}}(n_{i},
\gamma_{\alpha\beta})$ at which the localization length diverges,
$L_{\mathrm{loc}}(E\! \rightarrow\! E_{\mathrm{ME}})\rightarrow\infty$. At a
critical concentration of impurities depending on the scattering parameters,
$n_{cr}(\gamma_{\alpha\beta})$, the mobility edge reaches zero energy,
$E_{\mathrm{ME}}[n_{cr}(\gamma_{\alpha\beta}), \gamma_{\alpha\beta}]=0$, and
all states become delocalized.

In the calculations of the concentration dependences presented in the Figs.\
4a,b we took into account suppression of the average gap parameter described by
Eqs.\ (\ref{GapSCTM1}) and (\ref{GapSCTM2}) and corresponding increase of the
coherence length $\xi=v_F/\Delta$.

\section{Density of states at small concentration of impurities:
pairs-dominated regime}\label{app.pairs}

\begin{figure}[ptb]
\begin{center}
\includegraphics[width=3.2in]{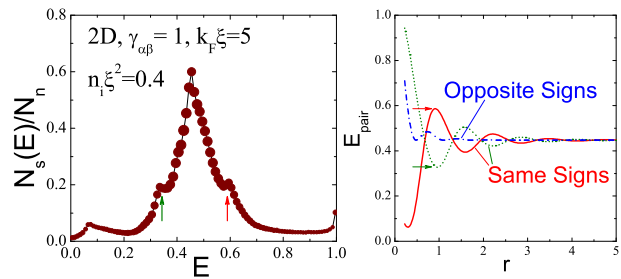}
\end{center}
\caption{(color online) \emph{Left:} The 2D DoS for small concentration of
impurities, same as in Fig.1 of the main paper. \emph{Right:} The energy of
impurity pair as a function of the distance between impurities for same-sign
and opposite-sign impurities. The arrows mark the extremum points which account
for the small peaks in the DoS.} \label{Fig-DoS-pairs}
\end{figure}
At small concentrations of impurities, $n_{i}\xi^{d}\ll1$, the density of state
is determined by impurity pairs.\cite{LifshitzPGBook} The interaction between
two close impurities gives the correction to the energy\cite{RusinovPZheTF68},
$E_{\mathrm{pair}}(r)=\varepsilon_0+\delta \varepsilon (r)$ and the behavior of
the correction $\delta \varepsilon (r)$ depends on the relative sign of the
off-diagonal scattering potential $U_{12}$ for two impurities. For same-sign
impurities the single-site energy level splits into two levels corresponding to
symmetric and antisymmetric combinations of the wave functions at the impurity
sites. The separation-dependent energy corrections, $\delta\varepsilon_{\pm}$,
rapidly oscillate with distance between the impurities $r$ as
\[
\delta\varepsilon_{\pm}(r)\propto\pm\frac{\sin(k_{F}r+\alpha_{d})}
{(k_{F}r)^{(d-1)/2}}\exp\left(  -\sqrt{1-\varepsilon_{0}^{2}}r/\xi\right),
\]
where $\xi=v_{F}/\Delta$ is the coherence length and $d$ is space
dimensionality, see Fig.\ \ref{Fig-DoS-pairs}. For opposite-sign impurities the
energy level remains double-degenerate with two states corresponding to
localization near two impurity sites. The separation-dependent energy shift in
this case is always positive and much smaller than for the same-sign
impurities,
\[
\delta\varepsilon(r)\propto\frac{\sin^{2}(k_{F}r+\alpha_{d})}{(k_{F}R)^{d-1}
}\exp\left(  -2\sqrt{1-\varepsilon_{0}^{2}}r/\xi\right).
\]
The coefficients in the energy corrections depend on the scattering parameters
of impurities.

The contribution to the DoS coming from impurity pairs with separation $r_{p}$
less than the typical distance, $r_{p}\ll n_{i}^{-d}$, can be evaluated in a
simple way.\cite{LifshitzPGBook} The concentration of the impurity pairs with
separations between $r_{p}$ and $r_{p}+dr_{p}$ is given by
$\frac{A_{d}}{2}n_{i}^{2} r_{p}^{d-1}dr_{p}$ with $A_{2}=2\pi$ and
$A_{3}=4\pi$. The contribution to the DoS at the energy $E$ is given by the
pairs satisfying the equation $E=E_{\mathrm{pair}} (r_p)$ and for a
nonmonotonic dependence $E_{\mathrm{pair}}(r)$, this equation may have several
solutions. The pair contribution to the DoS can be evaluated as
\[
N_{s}(E)=\frac{A_{d}}{2}n_{i}^{2}\sum_{r_p}\left[r_{p}(E)\right]
^{d-1}\left\vert \frac{dr_{p}}{dE}\right\vert ,
\]
where the sum is taken over all values of $r_p$ corresponding to the same
energy.

For pair separations corresponding to the energy extrema,
$E_{\mathrm{pait}}(r)\approx E_{e}+\frac{a} {2}(r-r_{e})^{2}$, the
isolated-pairs approximation gives a divergency at $E\rightarrow E_e$,
\[
N_{s}(E)\approx\frac{A_{d}n_{i}^{2}\left[  r_{e}\right]  ^{d-1}}
{\sqrt{2a\left(  E-E_{e}\right)  }}.
\]
These singularities, smeared by interactions with more remote impurities,
account for the small peaks found in the DoS at small concentrations, see Fig.
\ref{Fig-DoS-pairs}. The energy for opposite-sign impurities has a series of
minima at exactly $E=\varepsilon_{0}$ meaning that there is also such a pair
singularity at the peak center. This explains the sharpness of the peak at
small concentrations of impurities.

\end{document}